\documentclass[a4paper]{report}
\usepackage[utf8]{inputenc}
\usepackage[T1]{fontenc}
\usepackage{RJournal}
\usepackage{amsmath,amssymb,array}
\usepackage{booktabs}
\usepackage{nameref,hyperref}
\usepackage{amsfonts}
\usepackage{multirow}
\usepackage{amsthm,bbm,bm}
\usepackage{natbib}
\usepackage{pgfplots}

\def\T{{ \mathrm{\scriptscriptstyle T} }}
\def\bB{{ \mathbf{B} }}

\begin{document}

\sectionhead{Contributed research article}
\volume{XX}
\volnumber{YY}
\year{20ZZ}
\month{AAAA}

\begin{article}
\title{orthoDr: Semiparametric Dimension Reduction via Orthogonality Constrained Optimization}
\author{by Ruoqing Zhu, Jiyang Zhang, Ruilin Zhao, Peng Xu, Wenzhuo Zhou and Xin Zhang}

\maketitle

\abstract{
\CRANpkg{orthoDr} is a package in R that solves dimension reduction problems using orthogonality constrained optimization approach. The package serves as a unified framework for many regression and survival analysis dimension reduction models that utilize semiparametric estimating equations. The main computational machinery of \pkg{orthoDr} is a first-order algorithm developed by \cite{wen2013feasible} for optimization within the Stiefel manifold. We implement the algorithm through Rcpp and OpenMP for fast computation. In addition, we developed a general-purpose solver for such constrained problems with user-specified objective functions, which works as a drop-in version of optim(). The package also serves as a platform for future methodology developments along this line of work.
}

\section{Introduction}

Dimension reduction is a long-standing problem in statistics and data science. While the traditional principal component analysis \citep{jolliffe1986principal} and related works provide a way of reducing the dimension of the covariates, the term ``sufficient dimension reduction'' is more commonly referring to a series of regression works originated from the seminal paper on sliced inverse regression \citep{li1991sliced}. In such problems, we observe an outcome $Y \in \mathbb{R}$, along with a set of covariates $X = (X_1, \ldots, X_p)^\T \in \mathbb{R}^p$. Dimension reduction models are interested in modeling the conditional distribution of $Y$ given $X$, while their relationship satisfies, for some $p \times d$ matrix $\bB = (\boldsymbol\beta_1, \ldots, \boldsymbol\beta_p)$,
\begin{equation}
Y = h(X, \epsilon) = h(\bB^\T X, \epsilon) = h(\boldsymbol\beta_1^\T X, \ldots, \boldsymbol\beta_d^\T X, \epsilon), \label{dr:model}
\end{equation}
where $\epsilon$ represents any error terms and $h$, with a slight abuse of notation, represents the link function using $X$ or $\bB^\T X$. One can easily notice that when $d$, the number of columns in $\bB$, is less than $p$, a dimension reduction is achieved, in the sense that only a $d$ dimensional covariate information is necessary for fully describing the relationship \citep{cook2009regression}. Alternatively, this relationship can be represented as \citep{zeng2010integral}
\begin{equation}
Y \perp X \mid \bB^\T X, \label{dr:perp}
\end{equation}
which again describes the sufficiency of $\bB^\T X$. Following the work of \cite{li1991sliced}, a variety of methods have been proposed. An incomplete list of literature includes \cite{cook1991discussion, cook1999dimension, yin2002dimension, chiaromonte2002sufficient, zhu2006sliced, li2007directional, zhu2010dimension, zhu2010sufficient, cook2010envelope, lee2013general, cook2014fused, li2017parsimonious}. For a more comprehensive review of the literature, we refer the readers to \cite{ma2013review}. One advantage of many early developments in dimension reduction models is that only a singular value decomposition is required to obtain the reduced space parameters $\bB$ through inverse sliced averaging. However, this comes at a price of assuming the linearity assumption \citep{li1991sliced}, which is almost the same as assuming that the covariates follow an elliptical distribution \citep{li2009dimension, dong2010dimension}. Moreover, some methods require more restrictive assumptions on the covariance structure \citep{cook1991discussion}. Many methods attempt to avoid these assumptions by resorting to nonparametric estimations. The most successful ones include \cite{xia2002adaptive} and \cite{xia2007constructive}. However, recently a new line of work started by \cite{ma2012semiparametric, ma2012efficiency, ma2013efficient} shows that by formulating the problem into semiparametric estimating equations, not only we can avoid many distributional assumptions on the covariates, the obtained estimator of $\bB$ also enjoys efficiency. Extending this idea, \cite{sun2017counting} developed a framework for dimension reduction in survival analysis using a counting process based estimating equations. The method performs significantly better than existing dimension reduction methods for censored data such as \cite{li1999dimension, xia2010dimension} and \cite{lu2011sufficient}. Another recent development that also utilizes this semiparametric formulation is \cite{zhao2017efficient}, in which an efficient estimator is derived.

Although there are celebrated theoretical and methodological advances, estimating $\bB$ through the semiparametric estimating equations is still not a trivial task. Two challenges remain: first, by a careful look at the model definition \ref{dr:model}, we quickly noticed that the parameters are not identifiable unless certain constraints are placed. In fact, if we let $\mathbf{A}$ be any $d \times d$ full rank matrix, then $(\bB\mathbf{A})^\T X$ preserves the same column space information of $\bB^\T X$, hence, we can define $h^\ast((\bB\mathbf{A})^\T X, \epsilon)$ accordingly to retain exactly the same model as \eqref{dr:model}. While traditional methods can utilize singular value decompositions (SVD) of the estimation matrix to identify the column space of $\bB$ instead of recovering each parameter \citep{cook1999dimension}, it appears to be a difficult task in the semiparametric estimating equation framework. One challenge is that if we let $\bB$ change freely, the rank of the $\bB$ matrix cannot be guaranteed, which makes the formulation meaningless. Hence, for both computational and theoretical concerns, \cite{ma2012semiparametric} resorts to an approach that fixes the upper $d \times d$ block of $\bB$ as an identity matrix, i.e., $\bB = (\mathbf{I}_{d \times d}, \,\bB^{\ast\T})^\T$, where $\bB^\ast$ is a $(p-d) \times d$ matrix that sits in the lower block of $\bB$. Hence, in this formulation, only $\bB^\ast$ needs to be solved. While the solution is guaranteed to be rank $d$ in this formulation, as pointed out by \cite{sun2017counting}, this approach still requires correctly identifying and reordering of the covariate vector $\mathbf{x}$ such that the first $d$ entries are indeed important, which creates another daunting task. Another challenge is that solving semiparametric estimating equations requires the estimation of nonparametric components. These components need to be computed through kernel estimations, usually the Nadaraya-Watson type, which significantly increases the computational intensity of the method considering that these components need to be recalculated at each iteration of the optimization. Up to date, these drawbacks remain as the strongest criticism of the semiparametric approaches. Hence, although enjoying superior statistical asymptotic properties, are not as attractive as a traditional sliced inverse type of approaches such as \cite{li1991sliced} and \cite{cook1991discussion}.

The goal of our \CRANpkg{orthoDr} package is to develop a computationally efficient optimization platform for solving the semiparametric estimating equation approaches proposed in \cite{ma2013efficient}, \cite{sun2017counting} and possibly any future work along this line. Revisiting the rank preserving problem of $\bB$ mentioned above, we can essentially set a constraint that
\begin{equation}\label{B:constraint}
\bB^\T \bB = \mathbf{I},
\end{equation}
where $\mathbf{I}$ is a $d \times d$ identity matrix. A solution of the estimating equations that satisfies the constraint will correctly identify the dimensionality-reduced subspace. This is known as optimizing on the Stiefel manifold, which is a class of well-studied problems \citep{edelman1998geometry}. A recent R development \citep{martin2016manifoldoptim} utilizes quasi-Newton methods such as the well known BFGS method on the Riemannian manifold \citep{huang2016roptlib}. However, Second order optimization methods always require forming and storing large hessian matrices. In addition, they may not be easily adapted to penalized optimization problems, which often appear in high dimensional statistical problems \cite{zhu2006sliced, li2008sliced}. On the other hand, first-order optimization methods are faster in each iteration, and may also incorporate penalization in a more convenient way \cite{wen2010alternating}. By utilizing the techniques developed by \cite{wen2013feasible}, we can effectively search for the solution in the Stiefel manifold, and this becomes the main machinery of our package. Further incorporating the popular \CRANpkg{Rcpp} \citep{eddelbuettel2011Rcpp} and \CRANpkg{RcppArmadillo} \citep{eddelbuettel2014RcppArmadillo} toolboxes and the OpenMP parallel commuting, the computational time for our package is comparable to state-of-the-art existing implementations (such as \CRANpkg{ManifoldOpthm}), making the semiparametric dimension reduction models more accessible in practice.

The purpose of this article is to provide a general overview of the \pkg{orthoDr} package (version 0.6.2) and provide some concrete examples to demonstrate its advantages. \pkg{orthoDr} is available from the Comprehensive R Archive Network (CRAN) at \url{https://CRAN.R-project.org/package=orthoDr} and GitHub at \url{https://github.com/teazrq/orthoDr}. We begin by explaining the underlying formulation of the estimating equation problem and the parameter updating scheme that preserves orthogonality. Next, the software is introduced in detail using simulated data and real data as examples. We further demonstrate an example that utilizes the package as a general purpose solver. We also investigate the computational time of the package compared with existing solvers. Future plans for extending the package to other dimension reduction problems are also discussed.

\section{Model description}

\subsection{Counting process based dimension reduction}
To give a concrete example of the estimating equations, we use the semiparametric inverse regression approach defined in \cite{sun2017counting} to demonstrate the calculation. Following the common notations in the survival analysis literature, let $X_i$ be the observed $p$ dimensional covariate values of subject $i$, $Y_i = \min(T_i, C_i)$ is the observed survival time, with failure time $T_i$ and censoring time $C_i$, and $\delta_i = I(T_i \leq C_i)$ is the censoring indicator. A set of i.i.d. observations $\{X_i, Y_i, \delta_i\}_{i=1}^n$ is observed. We are interested in a situation that the conditional distribution of failure time $T_i | X_i$ depends only on the reduced space $\bB^\T X_i$. Hence, to estimate $\bB$, the estimating equation is given by
\begin{equation}\label{eeq:IR-Semi}
\widehat{\psi}_n \big(\bB \big)\!=\!\text{vec} \Bigg[
\!\frac{1}{n}\sum_{i=1}^n \sum_{\substack{j=1 \\ \delta_j = 1}}^n \!\left\{ X_i \!-\! \widehat E \big(X \big| Y \geq Y_j, \bB^\T X_i\big) \right\} \widehat \varphi^\T(Y_j) \left\{\delta_i I(j \!=\! i) \!-\! \widehat \lambda\big(Y_j | \bB^\T X_i\big) \right\} \Bigg],
\end{equation}
where the operator $\text{vec}(\cdot)$ is the vectorization of matrix. Several components are estimated nonparasitically: the function $\widehat\varphi(u)$ is estimated by sliced averaging,
\begin{equation}\label{eq:phi}
\widehat \varphi(u)=\frac{\sum_{i=1}^n X_i I\big( u \leq Y_i < u + \triangle u, \delta_i = 1\big) }{\sum_{i=1}^n I\big( u \leq Y_i < u+\triangle u, \delta_i = 1\big)} - \frac{\sum_{i=1}^n X_i I\big(Y_i \geq u\big)}{\sum_{i=1}^n I\big(Y_i \geq u\big)},
\end{equation}
where $\triangle u$ is chosen such that there are $hn$ number of observations lie between $u$ and $u + \triangle u$. The conditional mean function $\widehat E \big(X | Y \geq u, \bB^\T X \! = \! z\big)\!$ is estimated through the Nadaraya-Watson kernel estimator
\begin{equation}\label{eq:condX}
\widehat E \big(X | Y \geq u, \bB^\T X \! = \! z\big)\! = \frac{\sum_{i=1}^n X_i K_h(\bB^\T X_i \!-\! z)I(Y_i \geq u) }{\textstyle\sum_{i=1}^n K_h(\bB^\T X_i \!-\! z) I(Y_i \geq u)}.
\end{equation}
In addition, the the conditional hazard function at any time point $u$ can be estimated by
\begin{equation}
\widehat \lambda(u | \bB^\T X = z) = \frac{\sum_{i=1}^n K_b(Y_i-u) \delta_i K_h\big(\bB^\T X_i - z\big)}{\sum_{j=1}^n I\big(Y_j \geq u\big) K_h \big(\bB^\T X_j - z\big)}.\label{eq:chf}
\end{equation}

However, this substantially increase the computational burden since the double kernel estimator requires ${\cal O}(n^2)$ flops to calculate the hazard at any given $u$ and $z$. Instead, an alternative version using \cite{dabrowska1989uniform} can greatly reduce the computational cost without compromising the performance. Hence, we estimate the conditional hazard function by
\begin{equation}\label{eq:lambda}
\widehat \lambda(u | \bB^\T X = z) = \frac{\sum_{i=1}^n I\big(Y_i = u\big)I\big(\delta_i = 1\big)K_h\big(\bB^\T X_i - z\big)}{\sum_{i=1}^n I\big(Y_i \geq u\big) K_h\big(\bB^\T X_i - z\big)},
\end{equation}
which requires only ${\cal O}(n)$ flops. In the above equations \eqref{eq:phi}, \eqref{eq:condX} and \eqref{eq:lambda}, $h$ is a pre-specified kernel bandwidth and $K_h(\cdot) = K(\cdot/h)/h$, where $K(\cdot)$ is the Gaussian kernel function. By utilizing the method of moments estimators \citep{hansen1982large} and noticing our constraint for identifying the column space of $\bB$, solving for the solution of the estimating equations \eqref{eeq:IR-Semi} is equivalent to
\begin{align}\label{eq:GMM}
\text{minimize} \quad & f(\bB) = \widehat{\psi}_n(\bB)^\T \widehat{\psi}_n(\bB)\\
\text{subject to} & \quad \bB^\T \bB = \mathbf{I}.
\end{align}
Essentially all other semiparametric dimension reduction models described in \cite{ma2013efficient}, and more recently \cite{ma2015validated} \cite{xu2016estimating}, \cite{sun2017counting}, \cite{huang2017effective} and many others can be estimated in the samimilar fashion as the above optimization problem. However, due to the difficult in the constrains and the purpose of identifiability, all of these methods resort to either fixing the upper block of the $\bB$ matrix as an identity matrix or adding a penalty of $\lVert \bB^\T \bB - \mathbf{I} \rVert_F$ to preserve the orthogonality constraint. There appears to be no existing method that solves \eqref{eq:GMM} directly. Here, we utilize \cite{wen2013feasible}'s approach which can effectively tackle this problem.

\subsection{Orthogonality preserving updating scheme}

The algorithm works in the same fashion as a regular gradient decent, except that we need to preserve the orthogonality at each iteration of the update. As described in \cite{wen2013feasible}, given any feasible point $\bB_0$, i.e., ${\bB_0}^\T \bB_0 = \mathbf{I}$, which can always be generated randomly, we update $\bB_0$ as follows. Let the $p \times d$ gradient matrix be
\begin{align}\label{eq:gradient}
\mathbf{G} = \left( \frac{\partial f(\bB_0)}{\partial \bB_0(i,j) } \right)_{\{i,j\}}.
\end{align}
Then, utilizing the Cayley transformation, we have
\begin{align}\label{eq:update}
\bB_{\text{new}} &= \Big(\mathbf{I} + \frac{\tau}{2} \mathbf{A} \Big)^{-1} \Big(\mathbf{I} - \frac{\tau}{2} \mathbf{A} \Big) \bB_0,
\end{align}
with the orthogonality preserving property $\bB_{\text{new}}^\T \bB_{\text{new}} = \mathbf{I}$. Here, $\mathbf{A} = \mathbf{G} {\bB_0}^\T - \bB_0 \mathbf{G}^\T$ is a skew-symmetric matrix. It can be shown that $\{\bB_{\text{new}}(\tau)\}_{\tau \geq 0}$ is a descent path. Similar to line search algorithms, we can then find a proper step size $\tau$ through a curvilinear search. Recursively updating the current value of $\bB$, the algorithm stops when the tolerance level is reached. An initial value is also important for the performance of nonconvex optimization problems. A convenient initial value for our framework is the computational efficient approach developed in \cite{sun2017counting}, which only requires a SVD of the estimation matrix.

\section{The R package orthoDr}

There are several main functions in the \pkg{orthoDr} package: \code{orthoDr\_surv}, \code{ortho\_reg} and \code{ortho\_optim}. They are corresponding to the survival model described perviously \citep{sun2017counting}, the regression model in \cite{ma2012semiparametric}, and a general constrained optimization function, respectively. In this section, we demonstrate the details of using these main functions, illustrate them with examples.

\subsection{Semiparametric dimension reduction models for survival data}\label{sec:surv}

The \code{orthoDr\_surv} function implements the optimization problem defined in Equation \eqref{eq:GMM}, where the kernel estimations and various quantities are implemented and calculated within C++. Note that in addition, the method defined previously, some simplified versions are also implemented such as the counting process inverse regression models and the forward regression models, which are all described in \cite{sun2017counting}. These specifications can be made using the \code{method} parameter. A routine call of the function \code{orthoDr\_surv} proceed as

\begin{example}
    orthoDr_surv(x, y, censor, method, ndr, B.initial, bw, keep.data,
                 control, maxitr, verbose, ncore)
\end{example}

\begin{itemize}
\item \code{x}: A matrix or data.frame for features (numerical only).
\item \code{y}: A vector of observed survival times.
\item \code{censor}: A vector of censoring indicators.
\item \code{method}: The estimating equation method used.
\begin{itemize}
    \item \code{"dm"} (default): semiparametric inverse regression given in \eqref{eeq:IR-Semi}.
    \item \code{"dn"}: counting process inverse regression.
    \item \code{"forward"}: forward regression model with one structural dimensional.
\end{itemize}
\item \code{ndr}: The number of structural dimensional. For \code{method} = \code{"dn"} or \code{"dm"}, the default is 2. For \code{method} = \code{"forward"} only one structural dimension is allowed, hence the parameter is suppressed.
\item \code{B.initial}: Initial \code{B} values. Unless specifically interested, this should be left as default, which uses the computational efficient approach (with the \code{CPSIR()} function) in \cite{sun2017counting} as the initial. If specified, must be a matrix with \code{ncol(x)} rows and \code{ndr} columns. The matrix will be processed by Gram-Schmidt if it does not satisfy the orthogonality constrain.
\item \code{bw}: A kernel bandwidth, assuming each variables have unit variance. By default we use the Silverman rule-of-thumb formula \cite{silverman2018density} to determine the bandwidth $$\text{\code{bw}} = 1.06\times\left(\frac4{d+2}\right)^\frac1{d+4}n^{-\frac1{d+4}}.$$
    This bandwidth can be computed using the \code{silverman(n, d)} function in our package.
\item \code{keep.data}: Should the original data be kept for prediction? Default is \code{FALSE}.
\item \code{control}: A list of tuning variables for optimization, including the convergence criteria. In particular, \code{epsilon} is the size for numerically approximating the gradient, \code{ftol}, \code{gtol}, and \code{btol} are tolerance levels for the objective function, gradients, and the parameter estimations, respectively, for judging the convergence. The default values are selected based on \cite{wen2013feasible} .
\item \code{maxitr}: Maximum number of iterations. Default is 500.
\item \code{verbose}: Should information be displayed? Default is \code{FALSE}.
\item \code{ncore}: Number of cores for parallel computing when approximating the gradients numerically. The default is the maximum number of threads.
\end{itemize}

We demonstrate the usage of \code{orthoDr\_surv} function by solving a problem with generated survival data.
\begin{example}
    # generate some survival data with two structural dimensions
    R> set.seed(1)
    R> N = 350; P = 6; dataX = matrix(rnorm(N*P), N, P)
    R> failEDR = as.matrix(cbind(c(1, 1, 0, 0, 0, 0, rep(0, P-6)),
     +  c(0, 0, 1, -1, 0, 0, rep(0, P-6))))
    R> censorEDR = as.matrix(c(0, 1, 0, 1, 1, 1, rep(0, P-6)))
    R> T = exp(-2.5 + dataX 
     +  failEDR[,1])*(dataX 
    R> C = exp( -0.5 + dataX 
    R> Y = pmin(T, C)
    R> Censor = (T < C)

    # fit the model
    R> orthoDr.fit = orthoDr_surv(dataX, Y, Censor, ndr = 2)
    R> orthoDr.fit

                 [,1]        [,2]
    [1,] -0.689222616  0.20206497
    [2,] -0.670750726  0.19909057
    [3,] -0.191817963 -0.66623300
    [4,]  0.192766630  0.68605407
    [5,]  0.005897188  0.02021414
    [6,]  0.032829356  0.06773089
\end{example}

To evaluate the accuracy of this estimation, a distance function \code{distance()} can be used. This function calculates the distance between the column spaces generated by the true $\bB$ and the estimated version $\widehat{\bB}$. Note that the canonical correlation distance is also closely related to the $\sin$ angle distance between the two column spaces.

\begin{example}
    distance(s1, s2, method, x)
\end{example}
\begin{itemize}
\item \code{s1}: A matrix for the first column space (e.g., $\bB$).
\item \code{s2}: A matrix for the second column space (e.g., $\widehat{\bB}$).
\item \code{method}:
\begin{itemize}
    \item \code{"dist"}: the Frobenius norm distance between the projection matrices of the two given matrices, where for any given matrix $\bB$, the projection matrix $\mathbf{P} = \bB(\bB^\T \bB)^{-1} \bB^\T$.
    \item \code{"trace"}: the trace correlation between two projection matrices $\textnormal{tr}(\mathbf{P} \widehat{\mathbf{P}})/d$, where $d$ is the number of columns of the given matrix.
    \item \code{"canonical"}: the canonical correlation between $\bB^\T X$ and $\widehat{\bB}^\T X$.
    \item \code{"sine"}: the sine angle distance $\|\sin{\Theta} \|_F$ obtained from $\mathbf{P}_1(\mathbf{I} - \mathbf{P}_2) = \mathbf{U} \sin{\Theta} \mathbf{V}^\T$.
\end{itemize}
\item \code{x}: The design matrix $X$ (default = \code{NULL}), required only if \code{method} $=$ \code{"canonical"} is used.
\end{itemize}

We compare the accuracy of the estimations obtained by the \code{method =}\code{"dm"} and \code{"dn"}. Note that the \code{"dm"} method enjoys double robustness property of the estimating equations, hence the result is usually better.

\begin{example}
    # Calculate the distance to the true parameters
    R> distance(failEDR, orthoDr.fit$B, "dist")

    [1] 0.1142773
    \end{example}
\begin{example}
    # Compare with the counting process inverse regression model
    R> orthoDr.fit1 = orthoDr_surv(dataX, Y, Censor, method = "dn", ndr = 2)
    R> distance(failEDR, orthoDr.fit1$B, "dist")

    [1] 0.1631814
  \end{example}

\subsection{Semiparametric dimension reduction models for regression}

The \code{orthoDr\_reg } function implements the semiparametric dimension reduction methods proposed in \cite{ma2012semiparametric}. A routine call of the function \code{orthoDr\_reg} proceed as
\begin{example}
    orthoDr_reg(x, y, method, ndr, B.initial, bw, keep.data, control,
                maxitr, verbose, ncore)
\end{example}

\begin{itemize}
\item \code{x}: A matrix or data.frame for features (numerical only).
\item \code{y}: A vector of observed continuous outcome.
\item \code{method}: We currently implemented two methods: the semiparametric sliced inverse regression method (\code{"sir"}), and the semiparametric principal Hessian directions method (\code{"phd"}).
\begin{itemize}
    \item \code{"sir"}: semiparametric sliced inverse regression method solves the sample version of the estimating equation $$E\Big(\big[E(X|Y)-E\{E(X|Y)|\bB^\T X\}\big]\big[X-E(X|\bB^\T X)\big]^\T\Big) = 0$$
    \item \code{"phd"}: semiparametric principal Hessian directions method that estimates $\bB$ by solving the sample version of  $$E\big[\{Y-E(Y|\bB^\T X)\}\{XX^\T - E(XX^\T | \bB^\T X)\}\big] = 0 $$
\end{itemize}
\item \code{ndr}: The number of structural dimensional (default is 2).
\item \code{B.initial}: Initial \code{B} values. For each method, the initial values are taken from the corresponding traditional inverse regression approach using the \pkg{dr} package. The obtained matrix will be processed by Gram-Schmidt for orthogonality.
\item \code{bw}, \code{keep.data}, \code{control}, \code{maxitr}, \code{verbose} and \code{ncore} are exactly the same as those in the \code{orthoDr\_surv} function.
\end{itemize}

To demonstrate the usage of \code{orthoDr\_reg}, we consider the problem of dimension reduction by fitting a semi-PHD model proposed by \cite{ma2012semiparametric}.

\begin{example}
    R> set.seed(1)
    R> N = 100; P = 4; dataX = matrix(rnorm(N*P), N, P)
    R> Y = -1 + dataX[,1] + rnorm(N)
    R> orthoDr_reg(dataX, Y, ndr = 1, method = "phd")

    Subspace for regression model using phd approach:
                [,1]
    [1,]  0.99612339
    [2,]  0.06234337
    [3,] -0.04257601
    [4,] -0.04515279
\end{example}

\subsection{Parallelled gradient approximation through OpenMP}

The estimation equations of the dimension reduction problem in the survival and regression settings usually have a complicated form. Especially, multiple kernel estimations are involved, which results in difficulties in taking derivatives analytically. As an alternative, numerically approximated gradients are implemented using OpenMP. A comparison between a single core and multiple cores (4 cores) is given in the following example. Results from 20 independent simulation runes are summarized in Table \ref{tab:openmp}. The data generating procedure used in this example is the same as the survival data used in Section \ref{sec:surv}. All simulations are performed on an i7-4770K CPU.

\begin{example}
    R> t0 = Sys.time()
    R> dn.fit = orthoDr_surv(dataX, Y, Censor, method = "dn", ndr = ndr,
     + ncore = 4, control = list(ftol = 1e-6))
    R> Sys.time() - t0
\end{example}

\begin{table}[h]
\centering
\caption{Computational cost of different numbers of cores}
\label{tab:openmp}
\smallskip
\begin{tabular}{lrr}
\hline
            & \multicolumn{2}{c}{\# of cores} \\ \cline{2-3}
            & \multicolumn{1}{c}{1} & \multicolumn{1}{c}{4} \\ \hline
$n\!=\!350$, $p\!=\!6$ & 3.9831  & 1.2741  \\
$n\!=\!350$, $p\!=\!12$ & 12.7780 & 3.4850  \\ \hline
\end{tabular}
\end{table}

\subsection{General solver for orthogonality constrained optimization}

\code{ortho\_optim} is a general purpose optimization function that can incorporate any user defined objective function $f$ (and gradient function if supplied). The usage of \code{ortho\_optim} is similar to the widely used \code{optim()} function. A routine call of the function proceed as
\begin{example}
  ortho_optim(B, fn, grad, ..., maximize, control, maxitr, verbose)
\end{example}

\begin{itemize}
\item \code{B}: Initial \code{B} values. Must be a matrix, and the columns are subject to the orthogonality constrains. It will be processed by Gram-Schmidt if not orthogonal.
\item \code{fn}: A function that calculates the objective function value. The first argument should be \code{B}. Returns a single value.
\item \code{grad}: A function that calculate the gradient. The first argument should be \code{B}. Returns a matrix with the same dimension as \code{B}. If not specified, a numerical approximation is used.
\item \code{...}: Arguments passed to \code{fn} and \code{grad} besides \code{B}.
\item \code{maximize}: By default, the solver will try to minimize the objective function unless \code{maximize = TRUE}.
\item The parameters \code{maxitr}, \code{verbose} and \code{ncore} works in the same way as introduced in the previous sections.
\end{itemize}

To demonstrate the simple usage of \code{ortho\_optim} as a drop-in function of \code{optim()}, we consider the problem of searching for the first principle component for a data matrix.

\begin{example}
    # an example of searching for the first principal component
    R> set.seed(1)
    R> N = 400; P = 100; X = scale(matrix(rnorm(N*P), N, P), scale = FALSE)
    R> w = gramSchmidt(matrix(rnorm(P), P, 1))$Q
    R> fx <- function(w, X) t(w) 
    R> gx <- function(w, X) 2*t(X) 

    # fit the model
    R> fit = ortho_optim(w, fx, gx, X = X, maximize = TRUE, verbose = 0)
    R> head(fit$B)

                [,1]
    [1,]  0.01268226
    [2,] -0.09065592
    [3,] -0.01471700
    [4,]  0.10583958
    [5,] -0.02656409
    [6,] -0.04186199

    # compare results with the prcomp() function
    R> library(pracma)
    R> distance(fit$B, as.matrix(prcomp(X)$rotation[, 1]), type = "dist")

    [1] 1.417268e-05
\end{example}

The \pkg{ManifoldOptim} \citep{martin2016manifoldoptim} package is known for solving optimization problems on manifolds. We consider the problem of optimizing Brockett cost function \citep{huang2016roptlib} on the Stiefel manifold with objective and gradient functions written in R. The problem can be stated as

\begin{equation}
\min_{\bB^\T \bB = \mathbf{I}_{p}, \, \, \bB\in \mathbb{R}^{n\times p}} \text{trace}(\bB^\T X \bB D),
\end{equation}

where $X \in \mathbb{R}^{n\times n}$, $X = X^\T $, $D = \text{diag}(\mu_1,\mu_2,...,\mu_p)$ with $\mu_1\geq\mu_2\geq...\geq\mu_p$. We generate the data with exactly the same procedure as the documentation file provided in the \pkg{ManifoldOptim} package, with only a change of notation. For our \pkg{orthoDr} package, the following code is used to specify the objective and gradient functions and solve for the optimal $\bB$.
\begin{example}
    R> n = 150; p = 5; set.seed(1)

    R> X <- matrix(rnorm(n*n), nrow=n)
    R> X <- X + t(X)
    R> D <- diag(p:1, p)

    R> f1 <- function(B, X, D) { Trace( t(B) 
    R> g1 <- function(B, X, D) { 2 * X 

    R> b1 = gramSchmidt(matrix(rnorm(n*p), nrow=n, ncol=p))$Q
    R> res2 = ortho_optim(b1, fn = f1, grad = g1, X, D)
    R> head(res2$B)

                 [,1]         [,2]         [,3]        [,4]         [,5]
    [1,] -0.110048632 -0.060656649 -0.001113691 -0.03451514 -0.063626067
    [2,] -0.035495670 -0.142148873 -0.011204859  0.01784039  0.129255824
    [3,]  0.052141162  0.015140614 -0.034893426  0.02600569  0.006868275
    [4,]  0.151239722 -0.008553174 -0.096884087  0.01398827  0.132756189
    [5,] -0.001144864 -0.056849007  0.080050182  0.23351751 -0.007219738
    [6,] -0.140444290 -0.112932425  0.082197835  0.18644089 -0.057003273
\end{example}

\begin{figure}[ht]
\centering
\caption{Log of function value vs. iteration ($n = 150$)}\label{fig:iter}
\includegraphics[width=2.65 in, height=2.5 in]{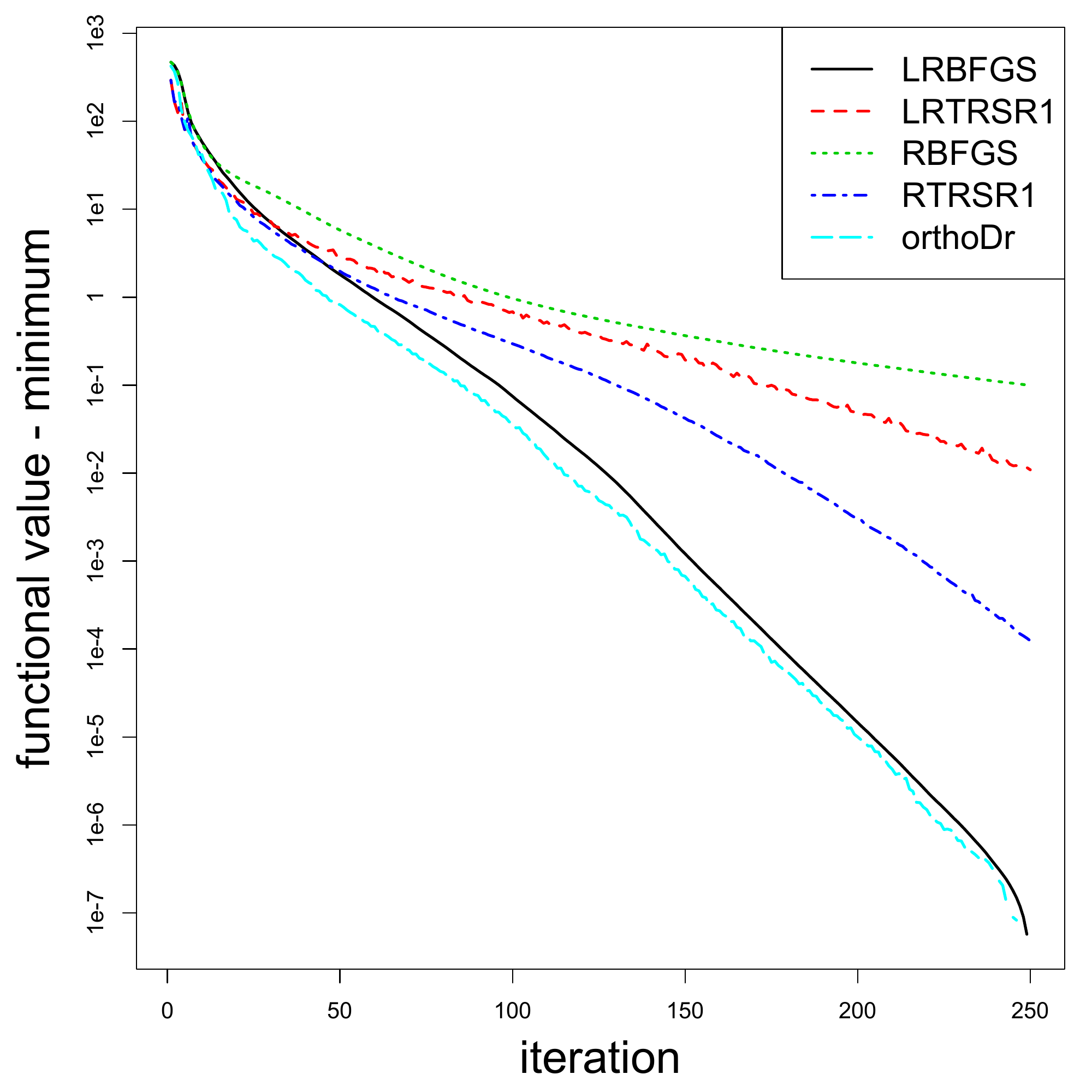}
\includegraphics[width=2.65 in, height=2.5 in]{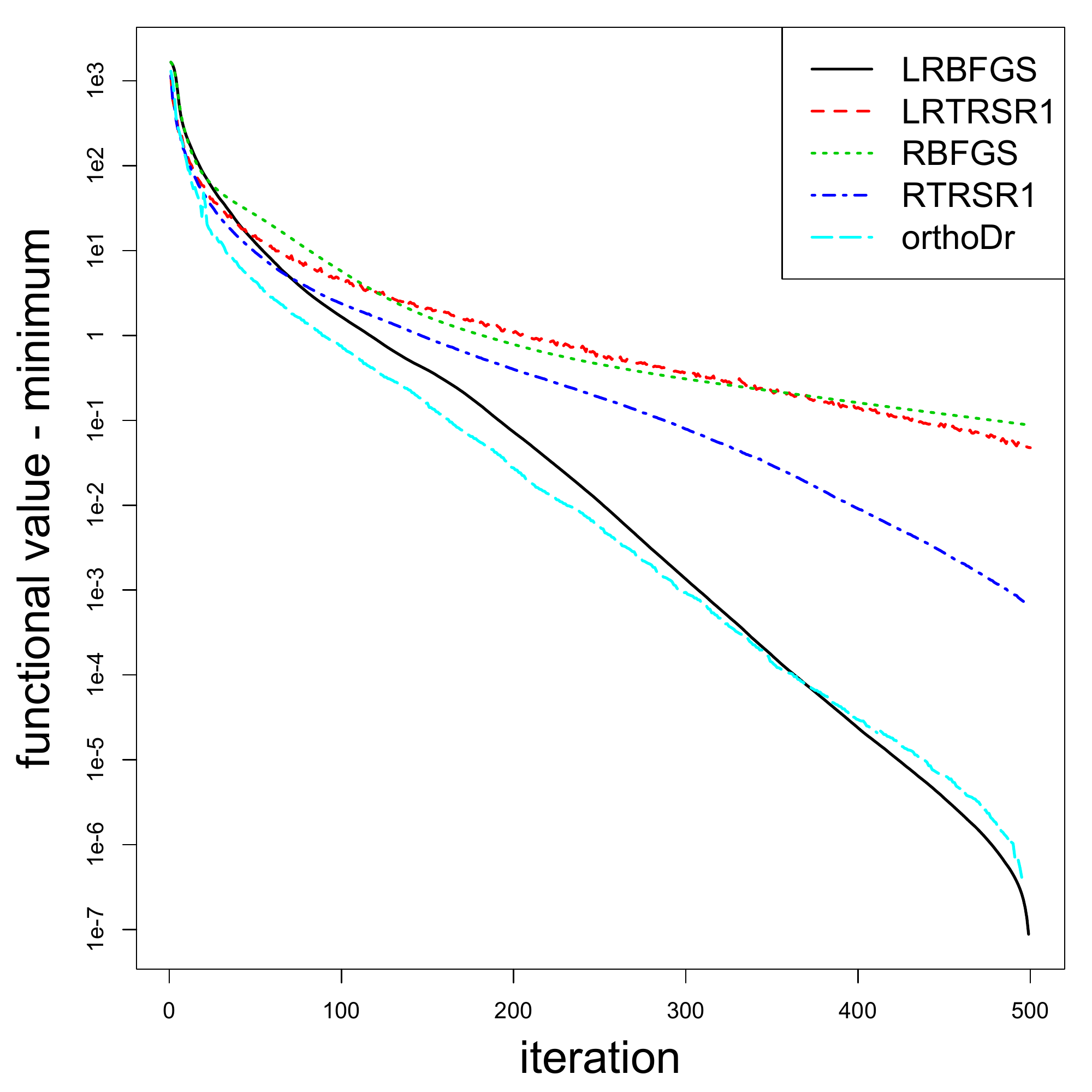}
\includegraphics[width=2.65 in, height=2.5 in]{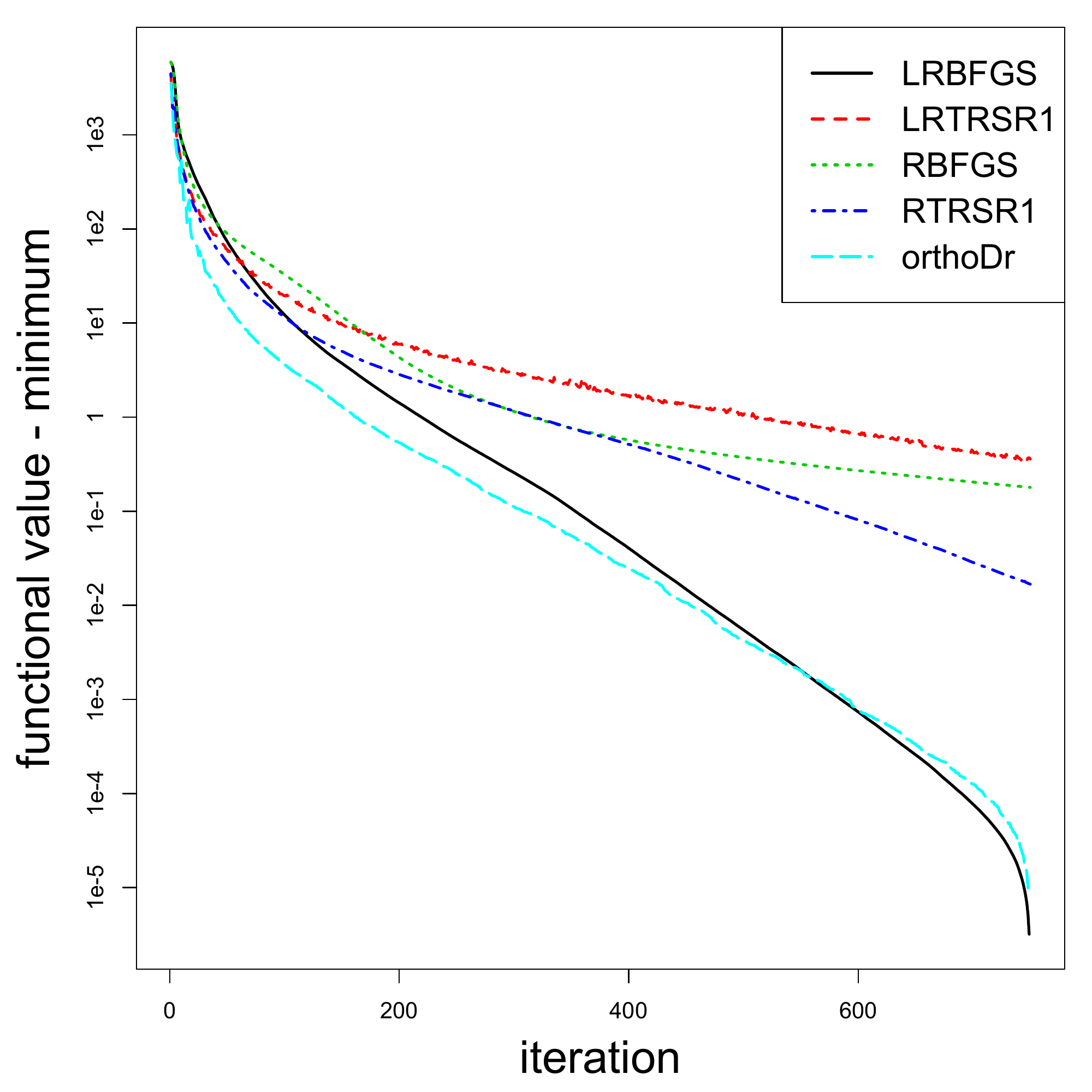}
\includegraphics[width=2.65 in, height=2.5 in]{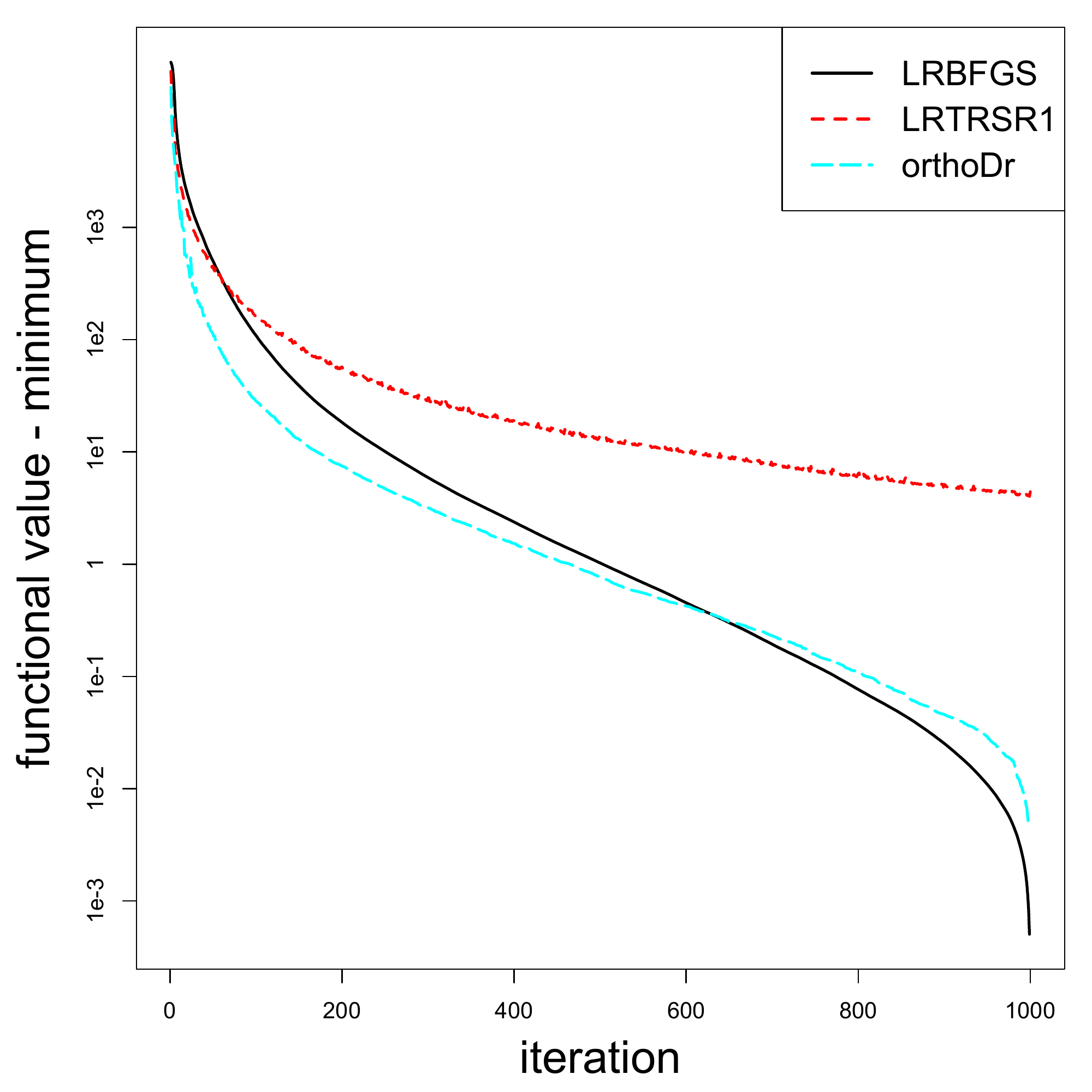}
From left to right, top to bottom: $p = 5, 10, 20$ and 50 respectively.
\end{figure}

\begin{figure}[ht]
\centering
\caption{Log of function value vs. iteration ($n = 500$)}\label{fig:iter2}
\includegraphics[width=2.65 in, height=2.5 in]{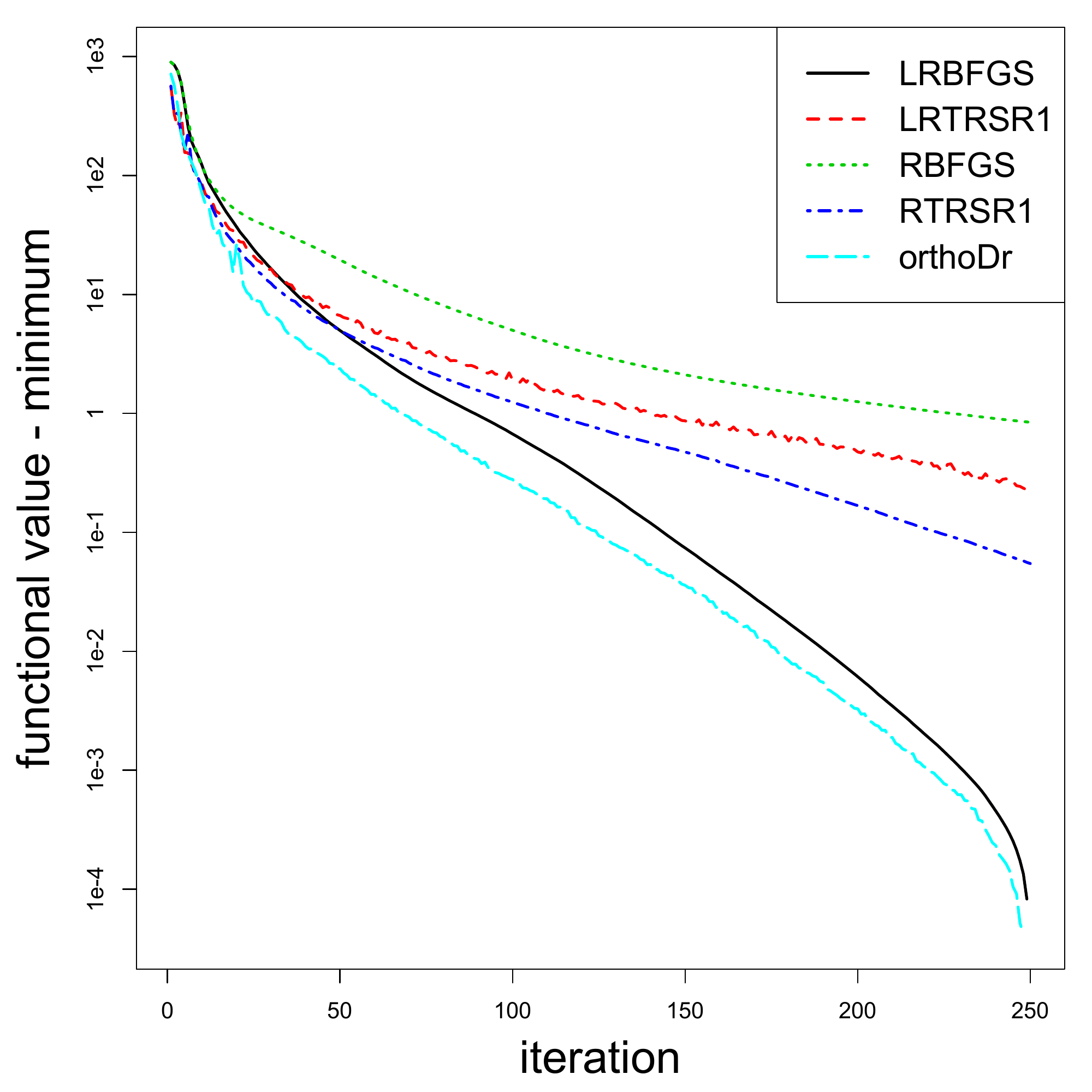}
\includegraphics[width=2.65 in, height=2.5 in]{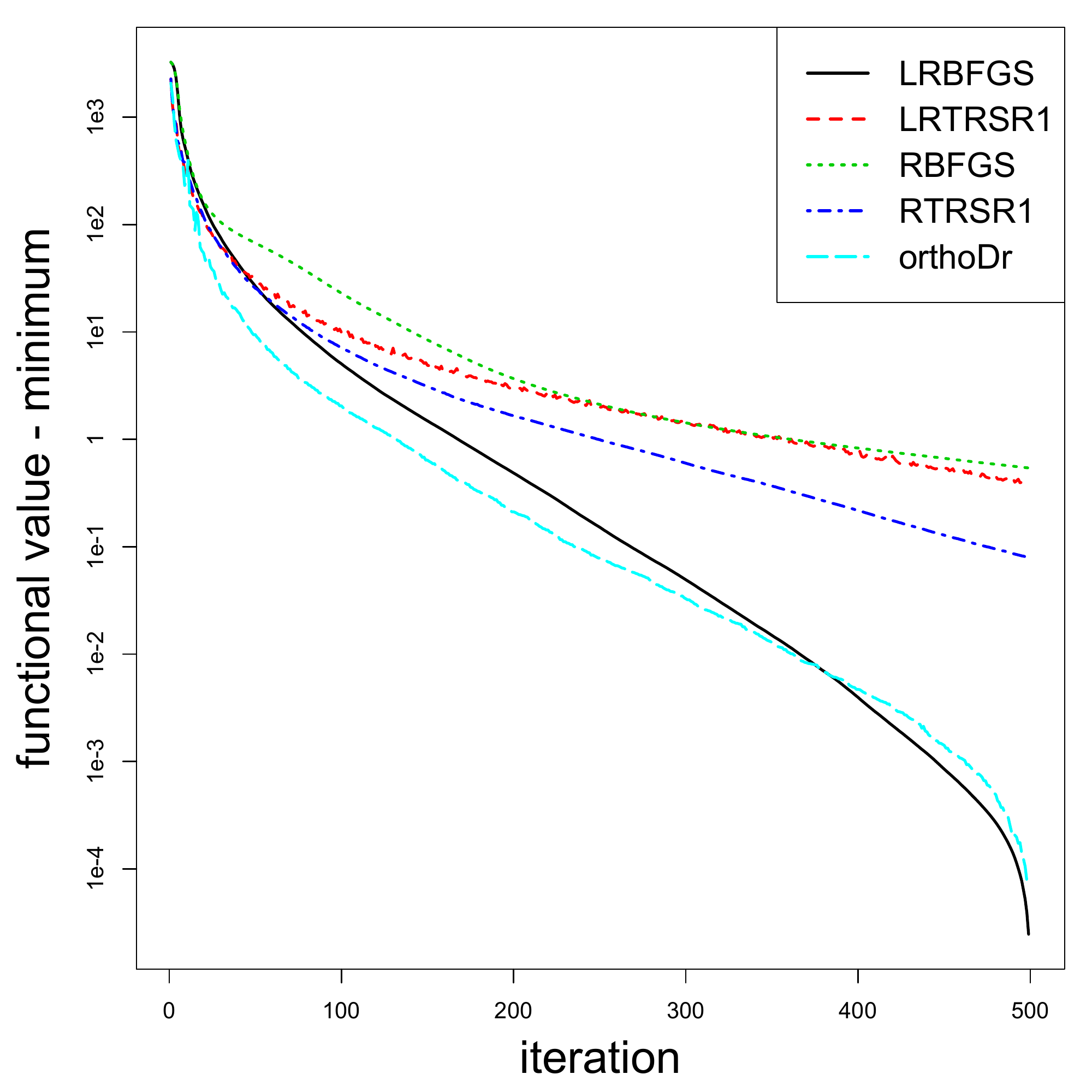}
\includegraphics[width=2.65 in, height=2.5 in]{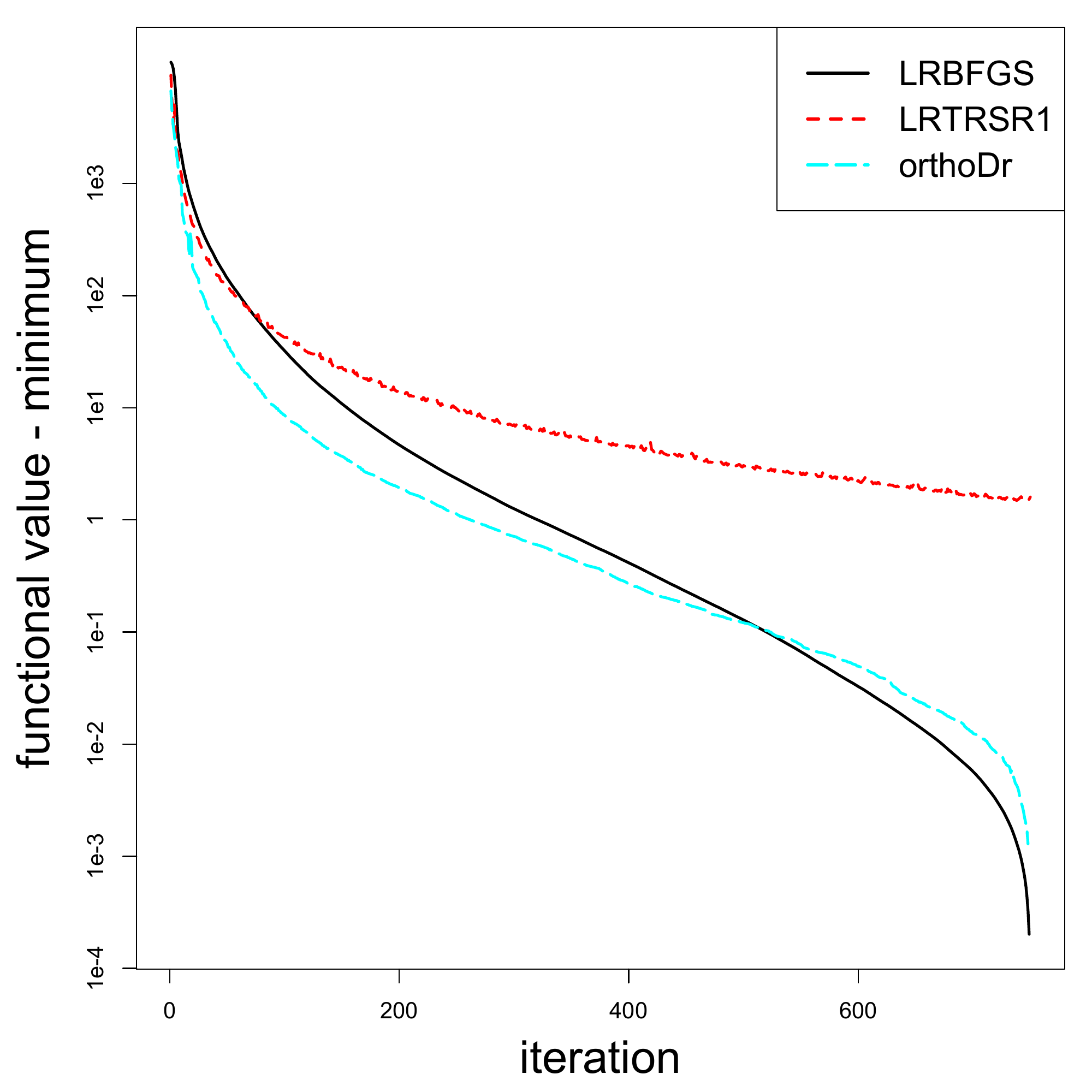}
\includegraphics[width=2.65 in, height=2.5 in]{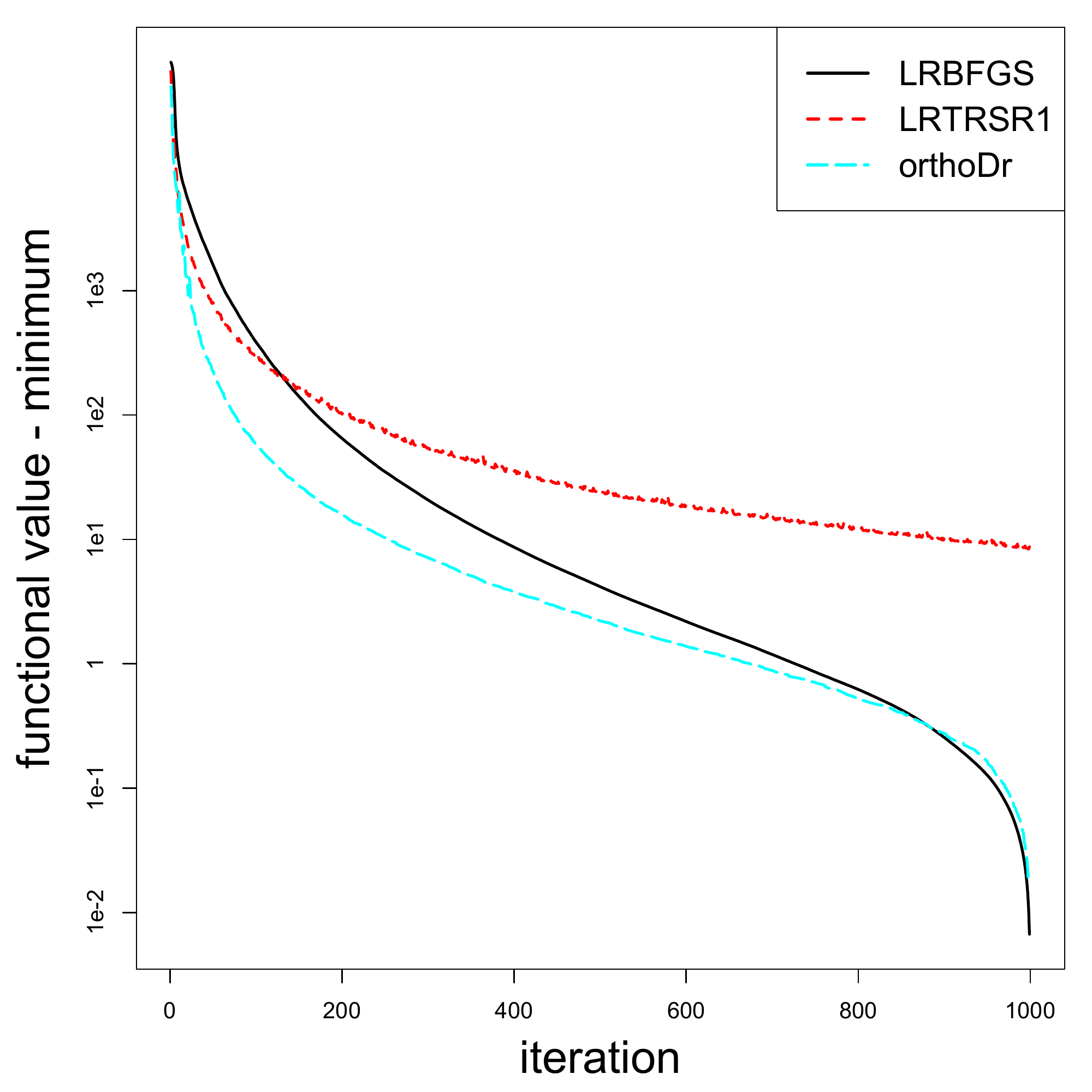}
From left to right, top to bottom: $p = 5, 10, 20$ and 50 respectively.
\end{figure}

Furthermore, we compare the performence with the \pkg{ManifoldOptim} package, using four optimization methods: \code{"LRBFGS"}, \code{"LRTRSR1"}, \code{"RBFGS"} and \code{"RTRSR1"} \citep{huang2016roptlib}. We wrote the same required functions for the Brockett problem in R. Further more, note that different algorithms implements slightly different stoping criterion, we run each algorithm a fixed number of iterations with a single core. We consider three smaller settings with $n = 150$, and $p = 5, 10$ and 15, and a larger setting with $n = 500$ and $p = 50$. Each simulation is repeated 100 times. The functional value progression (Figures \ref{fig:iter} and \ref{fig:iter2}) and the total time cost up to a certain number of iterations (Table \ref{tab:compare_time}) are presented.

We found that \code{"LRBFGS"} and our \pkg{orthoDr} package usually achieve the best performance, with functional value decreases the steepest in the log scale. In terms of computing time, \code{"LRBFGS"} and \pkg{orthoDr} performers similarly. Although \code{"LRTRSR1"} has similar computational time, its functional value falls behind. This is mainly because the theoretical complexity of second-order algorithms is similar to first order algorithms, both are of order ${\cal O}(p^3)$. However, it should be noted that for a semiparametric dimension reduction method, the major computational cost is not due to the parameter updates, rather, it is calculating the gradient since complicated kernel estimations are involved. Hence, we believe there is no significant advantage using either \code{"LRBFGS"} or our \pkg{orthoDr} package regarding the efficiency of the algorithm. However, first order algorithms may have an advantage when developing methods for penalized high-dimensional models.

\begin{table}[h]
\centering
\caption{Running times with a fixed number of iterations (in seconds)}
\small
\label{tab:compare_time}
\begin{tabular}{lllrrrrr}
\hline
\multicolumn{1}{c}{\multirow{2}{*}{$n$}}  & \multicolumn{1}{c}{\multirow{2}{*}{$p$}} & \multicolumn{1}{c}{\multirow{2}{*}{iteration}} & \multicolumn{4}{c}{\pkg{ManifoldOpthm}}   & \multicolumn{1}{c}{\multirow{2}{*}{\pkg{orthoDr}}}   \\ \cline{4-7}
\multicolumn{1}{c}{} & \multicolumn{1}{c}{} & \multicolumn{1}{c}{} & LRBFGS & LRTRSR1 & RBFGS    & RTRSR1 & \multicolumn{1}{c}{}  \\ \hline
\noalign{\smallskip}
$150$    &    5    &    250    &    0.053    &    0.062    &    0.451    &    0.452    &    0.065    \\
$150$    &    10    &    500    &    0.176    &    0.201    &    4.985    &    5.638    &    0.221    \\
$150$    &    20    &    750    &    0.526    &    0.589    &    28.084    &    36.142    &    0.819    \\
$150$    &    50    &    1000    &    2.469    &    2.662    & --    &    --    &    6.929    \\
\noalign{\smallskip}
$500$    &    5    &    250    &    0.403    &    0.414    &    7.382    &    7.426    &    0.423    \\
$500$    &    10    &    500    &    1.234    &    1.305    &    57.047    &    67.738    &    1.332    \\
$500$    &    20    &    750    &    3.411    &    3.6    &    --    &    --    &    3.974    \\
$500$    &    50    &    1000    &    13.775    &    14.43    &    --    &    --    &    19.862    \\
\noalign{\smallskip}
\hline
\end{tabular}
\end{table}

\section{Examples}

We use the \textit{Concrete Compressive Strength} \citep{yeh1998modeling} dataset as an example to further demonstrate the \code{orthoDr\_reg} function and to visualize the results. The dataset is obtained from the UCI Machine Learning Repository.

Concrete is the most important material in civil engineering. The concrete compressive strength is a highly nonlinear function of age and ingredients. These ingredients include cement, blast furnace slag, fly ash, water, superplasticizer, coarse aggregate, and fine aggregate. In this dataset, we have $n=1030$ observation, 8 quantitative input variables, and 1 quantitative output variable. We present the estimated two directions for structural dimension and further plot the observed data in these two directions. A non-parametric kernel estimation surface is further included to approximate the mean concrete strength.

\begin{example}
    R> concrete_data = read.csv(choose.files())
    R> X = as.matrix(concrete_data[,1:8])
    R> colnames(X) = c("Cement", "Blast Furnace Slag", "Fly Ash", "Water",
                "Superplasticizer", "Coarse Aggregate", "Fine Aggregate", "Age")
    R> Y = as.matrix(concrete_data[,9])

    R> result = orthoDr_reg(X, Y, ndr = 2, method = "sir", maxitr = 1000,
     + keep.data = TRUE)
    R> rownames(result$B) = colnames(X)
    R> result$B

                             [,1]         [,2]
    Cement             0.08354280 -0.297899899
    Blast Furnace Slag 0.27563507  0.320304097
    Fly Ash            0.82665328 -0.468889856
    Water              0.20738201  0.460314093
    Superplasticizer   0.43496780  0.540733516
    Coarse Aggregate   0.01141892  0.011870495
    Fine Aggregate     0.02936740 -0.004718979
    Age                0.02220664 -0.290444936
\end{example}

\begin{figure}
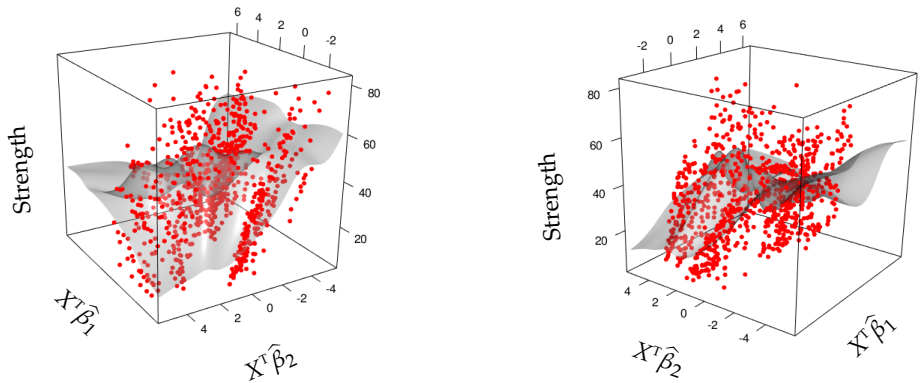

  \centering
\begin{tikzpicture}
\node[inner sep=0pt] (naive) at (0,0)
    {\includegraphics[width=2.4 in, height=2.3 in]{orthoDr_zhu_pic001.pdf}};
\node[text width=0.2 in, rotate=24] at (0.85, -2.7)
    {\small $ X^\T \widehat\beta_2$};
\node[text width=1 in, rotate=-43] at (-0.9, -2.6)
    {\small $ X^\T \widehat\beta_1$};
\node[text width=0.3 in, rotate=94] at (-2.3, -0.3)
    {\small Strength};
\end{tikzpicture}
\begin{tikzpicture}
\node[inner sep=0pt] (naive) at (0,0)
    {\includegraphics[width=2.4 in, height=2.3 in]{orthoDr_zhu_pic002.pdf}};
\node[text width=0.2 in, rotate=43] at (1.25, -2.2)
    {\small $ X^\T \widehat\beta_1$};
\node[text width=1 in, rotate=-20] at (-0.6, -2.75)
    {\small $ X^\T \widehat\beta_2$};
\node[text width=0.3 in, rotate=94] at (-2.85, -0.5)
    {\small Strength};
\node[text width=1 in, rotate=0] at (-3, -3.5)
    {\small };
\end{tikzpicture}
  \caption{Response variable over learned directions}\label{fig:concrete}
\end{figure}

\section{Discussion}
Using the algorithm proposed by \cite{wen2013feasible} for optimization on the Stiefel manifold, we developed the \pkg{orthoDr} package that serves specifically for semi-parametric dimension reductions problems. A variety of dimension reduction models are implemented for censored survival outcome and regression problems. In addition, we implemented parallel computing for numerically appropriate the gradient function. This is particularly useful for semi-parametric estimating equation methods because the objective function usually involves kernel estimations and the gradients are difficult to calculate. Our package can also be used as a general purpose solver and is comparable with existing manifold optimization approaches. However, since the performances of different optimization approaches could be problem dependent, hence, it could be interesting to investigate other choices such as the ``LRBFGS'' approach in the \pkg{ManifoldOptim} package.

Our package also serves as a platform for future methodology developments along this line of work. For example, we are currently developing a personalized dose-finding model with dimension reduction structure \citep{zhou2018dimension}. Also, when the number of covariates $p$ is large, the model can be over-parameterized. Hence, applying a $L_1$ penalty can force sparsity and allow the model to handle high-dimensional data. To this end, first-order optimization approaches can have advantages over second-order approaches. However, persevering the orthogonality during the Cayley transformation while also preserve the sparsity can be a challenging task and requires new methodologies. Furthermore, tuning parameters can be selected through a cross-validation approach, which can be implemented in the future.

\bibliography{orthoDr_zhu}

\address{Ruoqing Zhu\\
  Department of Statistics\\
  University of Illinois at Urbana-Champaign\\
  725 S. Wright St., 116 D\\
  Champaign, IL 61820, USA\\
  E-mail: \email{rqzhu@illinois.edu}\\
}

\address{Jiyang Zhang\\
  Department of Statistics\\
  University of Illinois at Urbana-Champaign\\
  725 S. Wright St.\\
  Champaign, IL 61820, USA\\
  E-mail: \email{jiyangz2@illinois.edu}\\
}

\address{Ruilin Zhao\\
  School of Engineering and Applied Science\\
  University of Pennsylvania\\
  220 South 33rd St.\\
  Philadelphia, PA 19104, USA\\
  E-mail: \email{rzhao15@seas.upenn.edu}\\
}

\address{Peng Xu\\
  Departments of Statistics\\
  Columbia University\\
  1255 Amsterdam Avenue\\
  New York, NY 10027, USA\\
  E-mail: \email{px2132@columbia.edu}\\
}

\address{Wenzhuo Zhou\\
  Department of Statistics\\
  University of Illinois at Urbana-Champaign\\
  725 S. Wright St.\\
  Champaign, IL 61820, USA\\
  E-mail: \email{wenzhuo3@illinois.edu}\\
}

\address{Xin Zhang\\
  Department of Statistics\\
  Florida State University\\
  214 OSB, 117 N. Woodward Ave.\\
  Tallahassee, FL 32306-4330, USA\\
  E-mail: \email{zhxnzx@gmail.com}\\
}

\end{article}

\end{document}